\documentclass[journal]{IEEEtran}

\usepackage{graphicx}
\usepackage{subfigure}

\begin{document}

\title{Process of Efficiently Parallelizing a Protein Structure Determination Algorithm}

\author{Michael~Bryson,
	Xijiang~Miao,
        Homayoun~Valafar% <-this % stops a space
	\thanks{Michael Bryson, Xijiang Miao, and Homayoun Valafar are part of the Bioinformatics Research Group at the University of South Carolina, Department of Computer Science.}% <-this % stops a space
}

\markboth{}{}

\maketitle

\begin{abstract}

Computational protein structure determination involves optimization in a problem space much too large to exhaustively search.  Existing approaches include optimization algorithms such as gradient descent and simulated annealing, but these typically only find local minima.  One novel approach implemented in {REDcRAFT} is to instead of folding a protein all at the same time, fold it residue by residue.  This simulates a protein folding as each residue exits from the generating ribosome.

While {REDcRAFT} exponentially reduces the problem space so it can be explored in polynomial time, it is still extremely computationally demanding.  This algorithm does have the advantage that most of the execution time is spent in inherently parallelizable code.  However, preliminary results from parallel execution indicate that approximately two thirds of execution time is dedicated to system overhead.  Additionally, by carefully analyzing and timing the structure of the program the major bottlenecks can be identified.  After addressing these issues, {REDcRAFT} becomes a scalable parallel application with nearly two orders of magnitude improvement.

\end{abstract}


\IEEEpeerreviewmaketitle

%%%%%%%%%%%%%%%%%%%%%%%%%%%%%%%%%%%%%%%%%%%%%%%%%%%%%%%%%%%%%%%%%%%%%%%%%%%%%%%%%%
%%%%%%%%%%%%%%%%%%%%%%%%%%%%%%%%%%%%%%%%%%%%%%%%%%%%%%%%%%%%%%%%%%%%%%%%%%%%%%%%%%

\section{Introduction}

\PARstart{S}{tructural} elucidation, including complete structures of individual domains of proteins as well as the assembly of biomolecular complexes, is often a requisite step in understanding fundamental physiological processes or in the design of drugs to combat disease. Therefore, the development of computational methods leading to rapid, cost-effective structure elucidation is an important task. In addition, it is important to develop methods that can simultaneously deal with internal motion in these assemblies.  Motion on a physiologically relevant time scale has always been suspected to play an important role in the biological function. The breathing motion of myoglobin \cite{biocomp06:Bertini, biocomp06:Cupane, biocomp06:Emerson, biocomp06:Shimada} can be cited as a historic example. Traditionally, full characterization of inter-molecular dynamics has been treated separately from structure elucidation, increasing the cost and time of these studies. Furthermore, conceptually, it is difficult to separate structure from dynamics since observables used for structure determination are perturbed by motion, and therefore any attempt at structure elucidation that disregards the dynamics (or vice versa) can produce faulty results \cite{biocomp06:DePristo}.  Here we present a method of simultaneous characterization of structure and dynamics of polypeptide chains from mainly the analysis of residual dipolar coupling data.  Residual dipolar couplings are observables obtained from Nuclear Magnetic Resonance (NMR) spectra that have come into use recently as a source of structural information \cite{biocomp06:Prestegard, biocomp06:Bax, biocomp06:Blackledge, biocomp06:Tjandra}.  The tool we present is unique in its simultaneous consideration of motion and structure based on inherent sensitivity of the used data to motions over the broad range of timescales spanning physiological processes.

Structure determination protocols based primarily on RDCs are becoming more common and require new programs that operate in fundamentally different ways from those that use NOE data.  Some of these have been put forward \cite{biocomp06:Jung, biocomp06:Rohl, biocomp06:Delaglio, biocomp06:Hus}. However, information richness and complexity of RDC data continues to challenge the abilities of any single analysis tool in existence today. Additionally, rapid development of new experimental methods of acquiring RDC data with continuously improving accuracy and precision, necessitate the existence of a parallel pursuit of information extraction methods. In this report, we present REDCRAFT ({REsidual Dipolar Coupling Residue Assembly and Filtering Tool}), a new open source analysis tool that accommodates the analysis of RDC data for simultaneous structure and dynamics characterization of proteins and polypeptides. Although utility of the current version of REDCRAFT is limited in application to proteins, the same principles can easily be applied to carbohydrates and nucleic acids.

%%%%%%%%%%%%%%%%%%%%%%%%%%%%%%%%%%%%%%%%%%%%%%%%%%%%%%%%%%%%%%%%%%%%%%%%%%%%%%%%%%
%%%%%%%%%%%%%%%%%%%%%%%%%%%%%%%%%%%%%%%%%%%%%%%%%%%%%%%%%%%%%%%%%%%%%%%%%%%%%%%%%%

\section{Background and Algorithm}

\subsection{Traditional Techniques}
Many approaches attempt to optimize \cite{biocomp06:Greshenfeld} the energy landscape \cite{biocomp06:Vasquez} in order to determine the most favorable structure.  Methods include gradient descent as implemented by Xplor-NIH \cite{biocomp06:Schwieters} as well as simulated annealing \cite{biocomp06:Greshenfeld}.  However, these algorithms usually suffer from a limited search space and local extrema.  The massive order of the problem therefore requires more innovative approaches outside of general optimization in order to more accurately determine a native structure.

%--------------------------------------------------------------------------------%

\subsection{REDcRAFT}
One promising algorithm is implemented as {REDcRAFT} \cite{biocomp06:ValafarMayer, biocomp06:Valafar}.  This program attempts to fold the strand sequentially instead of attacking the problem all at once.  It first considers all possible angles between each neighboring pair of residues.  Many of these are initially eliminated through the use of a Ramachandran \cite{biocomp06:Lovell, biocomp06:Ramachandran} filter and a scalar coupling filter \cite{biocomp06:Karplus}.  Residues are then added one at a time to the strand, each time considering only the most suitable combinations of previous angles.  By eliminating unfavorable combinations early in the algorithm, the search space of the problem is exponentially reduced.

\begin{figure}
\centering
\includegraphics[height=4.8in]{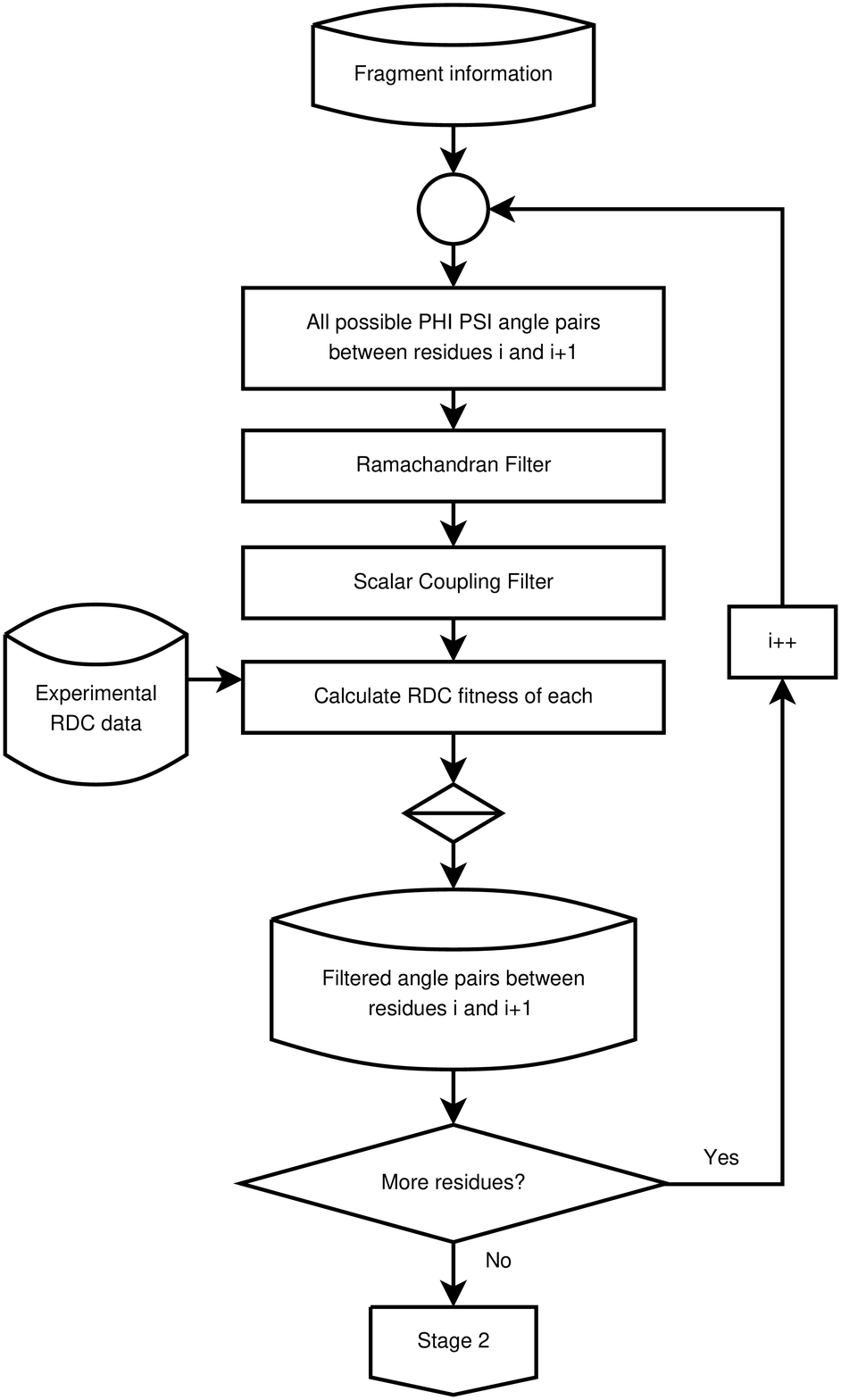}
\caption{Stage 1 of {REDcRAFT} first reads in information about the protein fragment.  It begins with the first pair of residues and generates a list of all possible $\phi$ and $\psi$ angles with resolution $R$.  This list is then reduced using Ramachandran and scalar coupling filters.  The fitness of each remaining angle pair is calculated by using experimental RDC data stored in a file.  The list is then sorted based on fitness and written to a file, and the process is repeated for every pair of neighboring residues.}
\label{fig:stage1}
\end{figure}

The program can more specifically be broken up into two stages.  The first stage is illustrated in Figure \ref{fig:stage1}.  The purpose of this stage is to take a protein sequence and generate a list of acceptable torsion angles by considering data from a pair of neighboring residues, based on the Ramachandran and scalar coupling filter.  These lists are then sorted based on how closely they mimic experimental Residual Dipolar Coupling (RDC) data.  These lists are stored as files, allowing trivial separation between the first and second stage of {REDcRAFT}.

\begin{figure}
\centering
\includegraphics[height=4.8in]{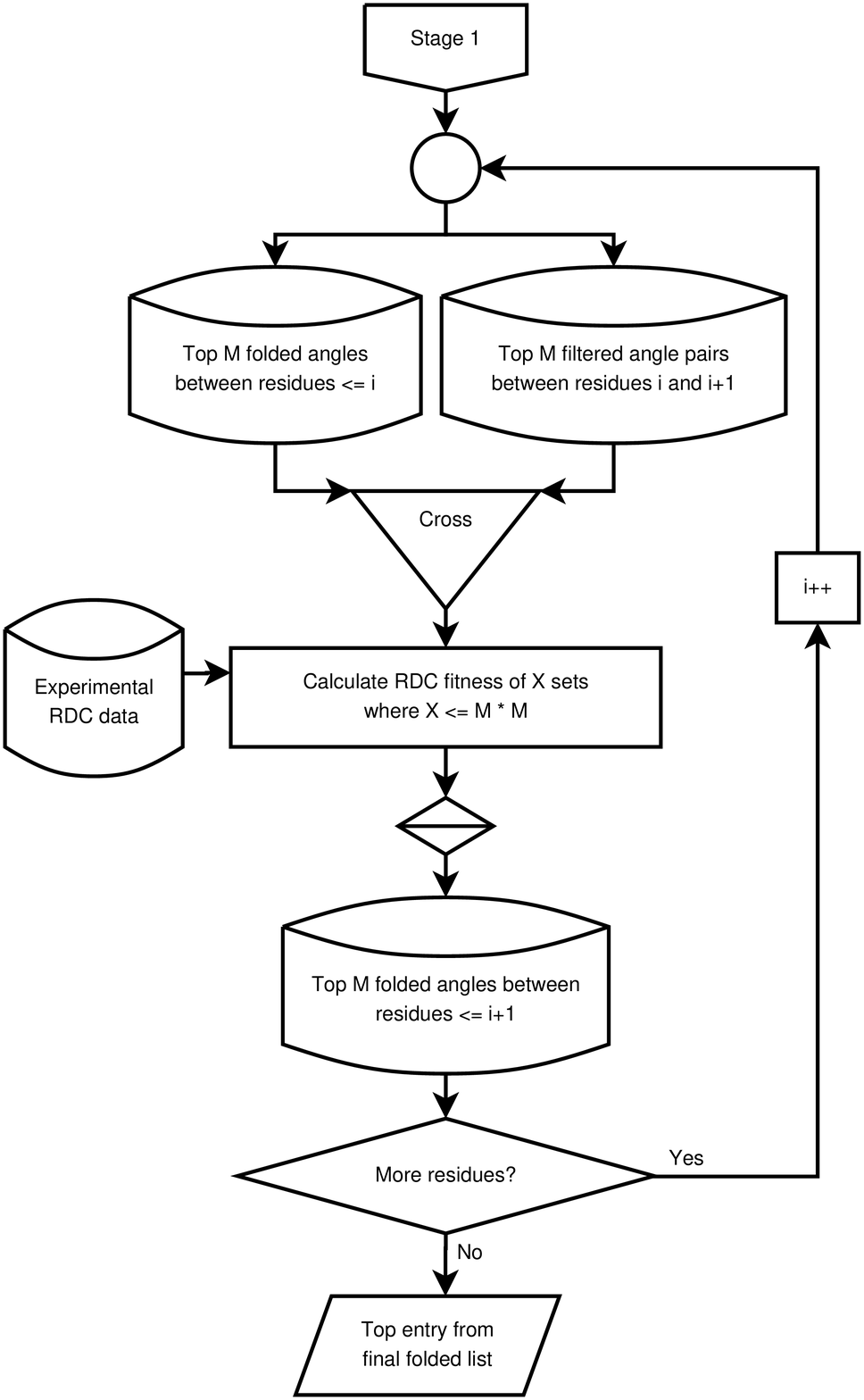}
\caption{Stage 2 of {REDcRAFT} begins by reading the angle lists generated in Stage 1 for the first two pairs of residues (two pairs of angles that construct the first three residues).  These two lists are then crossed to create every possible combination, and the fitness is calculated for each.  The combinations are sorted based on fitness and the top $M$ entries are written to a file, where $M$ is the search depth.  For the next iteration, this new file is read in along with the file containing the list of angle pairs between the last residue and the newly added residue.  This is repeated for the length of the protein fragment, and the final result is the top entry from the last sorted list.}
\label{fig:stage2}
\end{figure}

Once the lists have been generated, the second stage of {REDcRAFT} attempts to fold the fragment by adding one residue each iteration, as illustrated in Figure \ref{fig:stage2}. This is done by taking the $M$ most favorable sets of angles for the section of the fragment that has already been evaluated, and crossing it with the $M$ most favorable pairs of angles between the new residue and its attaching neighbor to create every possible combination.  Each of these $M^2$ angle combinations are then evaluated for fitness to experimental data and sorted.  This sorted list is then written back out as a file, to be used in the next iteration.

%--------------------------------------------------------------------------------%

\subsection{Search Space}
Considering every angle from -180 to 180 degrees with a resolution of 10 degrees gives 36 possible angles.  Each pair of residues has a $\phi$ and $\psi$ angle between them, giving $36 * 36$ possible combinations.  Therefore a protein with 50 residues would have $36^{98}$ possible combinations.

\begin{equation}
\label{eqn:total_configs}
C = (360 / R)^{2(N-1)}
\end{equation}

Equation \ref{eqn:total_configs} gives the number of possible combinations $C$ with angle resolution $R$ in degrees and number of residues $N$.  {REDcRAFT} however only considers the $M^2$ most favorable combinations at each iteration, so the total possible combinations with a search depth of $M$ is reduced to Equation \ref{eqn:total_configs_depth}.

\begin{equation}
\label{eqn:total_configs_depth}
C \leq (N - 1) M^2
\end{equation}

This algorithm therefore runs in polynomial time, yet can still be very computationally intensive.  An interesting note is that this runtime is not relative to angle resolution.  However a fine resolution with a small depth $M$ can lead to the rejection of possibly valid combinations, and a large resolution may not even consider valid combinations.  Therefore it is desirable to have as large of a search depth as computationally possible.

%%%%%%%%%%%%%%%%%%%%%%%%%%%%%%%%%%%%%%%%%%%%%%%%%%%%%%%%%%%%%%%%%%%%%%%%%%%%%%%%%%
%%%%%%%%%%%%%%%%%%%%%%%%%%%%%%%%%%%%%%%%%%%%%%%%%%%%%%%%%%%%%%%%%%%%%%%%%%%%%%%%%%

\section{Parallelization Strategy}

\subsection{Environment}

\begin{figure}
\centering
\includegraphics[width=1.8in]{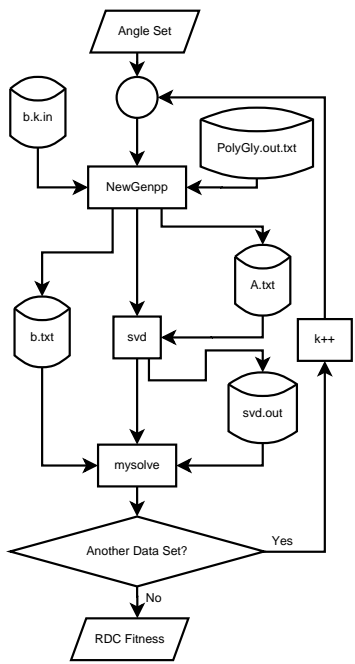}
\caption{Calculating RDC fitness makes use of three subprograms (NewGenpp, svd, and mysolve) that transfer data using files.  Two files are read only, while three additional files are created and read every iteration.  The angle set is passed in through the command line, and the fitness is passed back through standard output.  This process is repeated for each data set.}
\label{fig:calculate}
\end{figure}

The initial implementation of {REDcRAFT} was written with a mixture of C, C++, and Perl.  The core that calculates RDC fitness of each set of angles is illustrated in Figure \ref{fig:calculate}, where three separate programs (NewGenpp, svd, and mysolve) written in C/C++ are tied together using intermediate files.  This core is then wrapped with Perl scripts, one for each of Stage 1 and Stage 2.  There are also auxiliary Perl scripts used for data extraction and filter application.

Message Passing Interface (MPI) \cite{biocomp06:Gropp} was chosen for the parallel implementation due to its standardization, speed over queue based systems, and flexibility for future enhancements that could make use of communication between nodes.  Therefore the entire Stage 2 was implemented in C, with MPI used to scatter the angle combinations and gather the fitness calculations.  The computer system used for data collection is a 32 node Linux Beowulf cluster \cite{biocomp06:Becker}, where each node has a single 932 MHz Pentium III processor and 1 GB of memory.  During each run, processor usage is recorded at significant intervals of execution.

%--------------------------------------------------------------------------------%

\subsection{Algorithmic Analysis}
Before optimization begins, one needs to determine where the most time is spent in a program.  If the speed of a section of code that constitutes 5\% of a program's runtime is doubled, then a speedup of only 2.5\% has been achieved,  and the maximum theoretical speedup from this section could only be 5\%.  However if the speed of a section constituting 50\% of the program's runtime is doubled, a speedup of 25\% has been achieved.  Therefore it is crucial to determine where the most time is being spent, and why.

Since {REDcRAFT} consists of two easily separable stages, it is worthwhile to look at the individual runtimes.  Stage 1 has runtime $O(\frac{1}{R})$ relative to angle resolution $R$ for each residue, so given $N$ residues the total runtime is given by Equation \ref{eqn:complexity_stage1}.  Stage 2 however has runtime relative to search depth $M$.  Each iteration of Stage 2 has runtime $O(M^2)$, so given $N$ residues the total runtime is given by Equation \ref{eqn:complexity_stage2}.  Therefore given $M \gg \frac{360}{R}$, Stage 2 constitutes the majority of runtime.

\begin{equation}
\label{eqn:complexity_stage1}
O(\frac{N}{R})
\end{equation}

\begin{equation}
\label{eqn:complexity_stage2}
O(N * M^2)
\end{equation}

%--------------------------------------------------------------------------------%

\subsection{Parallel Algorithm}

\begin{figure*}
\centering
\includegraphics[width=5.5in]{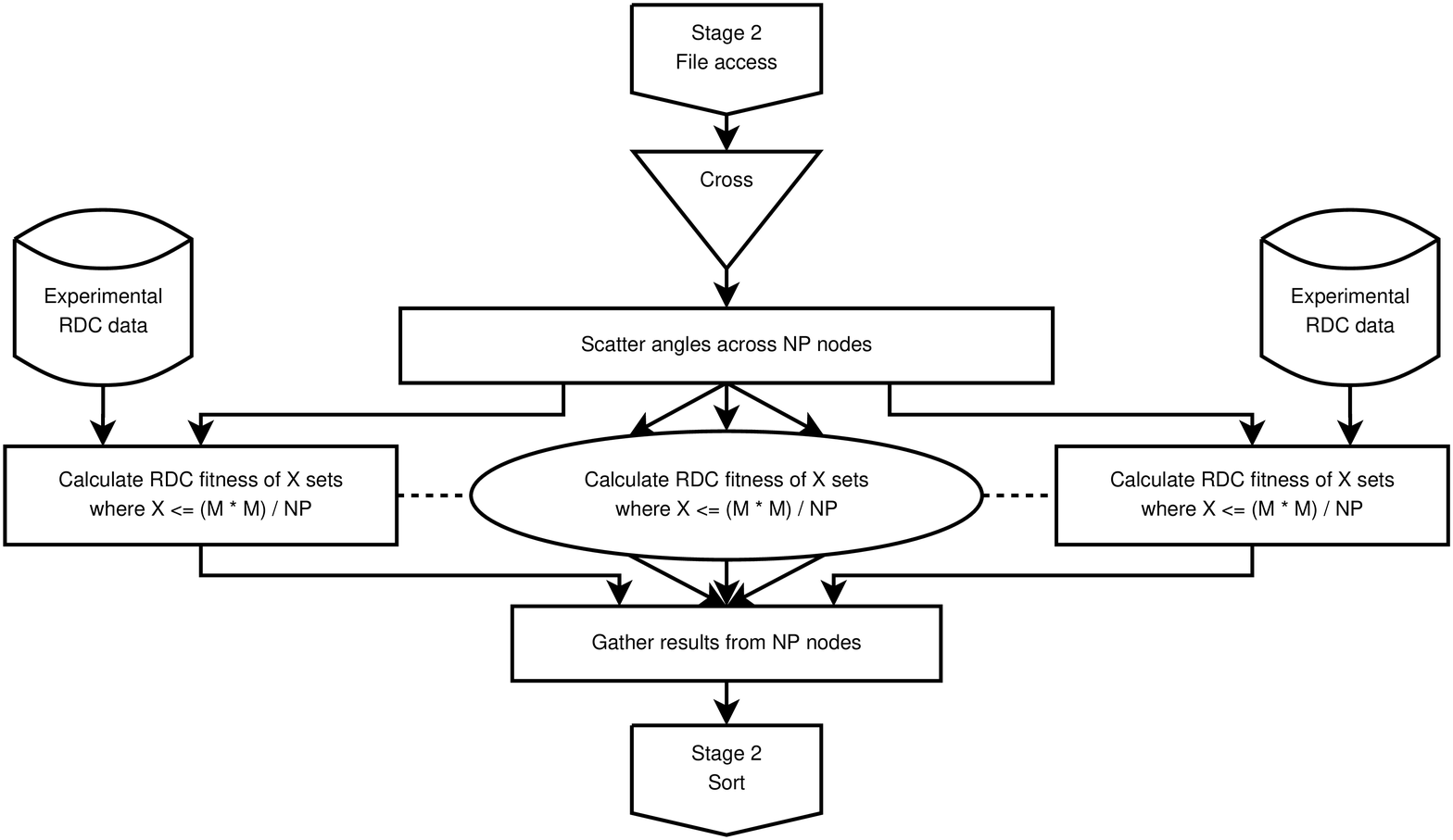}
\caption{The calculation section of Stage 2 can be parallelized by distributing the angle combinations to multiple processors.  After the angle lists are crossed, the resulting combinations are divided up and transferred to each of the $NP$ nodes to be used.  After all nodes are finished calculating their local combinations, the results are transferred back to the head node which procedes with the sorting step of Stage 2.  Note that scattering requires the transfer of a list of angle sets, while gathering only requires a list of single values to be transferred.  Therefore scattering is more costly than gathering, which can be seen in Figure \ref{fig:breakdown_noio_100}.}
\label{fig:parallelization}
\end{figure*}

Since Stage 1 of {REDcRAFT} only contributes to a small percentage of total runtime, optimization is focused on the second stage.  Of particular interest is the core calculation of RDC fitness which is naturally parallel, providing the potential for good scalability.  By distributing the list of $O(M^2)$ angle combinations across $NP$ nodes, the core calculation runtime can be reduced to $(\frac{SR}{NP} + OP)$ where $SR$ is the sequential runtime on a single node.  $OP$ is the parallelization overhead, such as time required for scattering and gathering \cite{biocomp06:Quinn} of data.  This parallel algorithm is illustrated in Figure \ref{fig:parallelization}.  Two questions arise from this however:  How much of the total runtime is spent in this core, and how large is $OP$?

%%%%%%%%%%%%%%%%%%%%%%%%%%%%%%%%%%%%%%%%%%%%%%%%%%%%%%%%%%%%%%%%%%%%%%%%%%%%%%%%%%
%%%%%%%%%%%%%%%%%%%%%%%%%%%%%%%%%%%%%%%%%%%%%%%%%%%%%%%%%%%%%%%%%%%%%%%%%%%%%%%%%%

\section{Implementation and Optimization}

\subsection{Initial Analysis}

\begin{figure*}
\centerline{
	\subfigure[$M = 25$]{
		\includegraphics[width=2.3in]{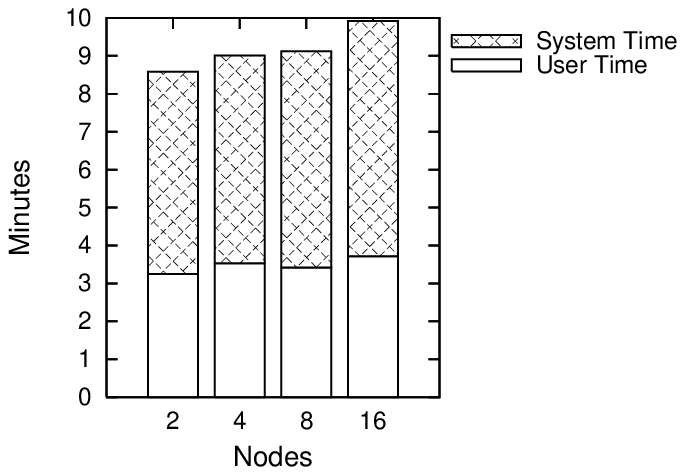}
		\label{fig:user_system_25}
	}
	\hfil
	\subfigure[$M = 50$]{
		\includegraphics[width=2.3in]{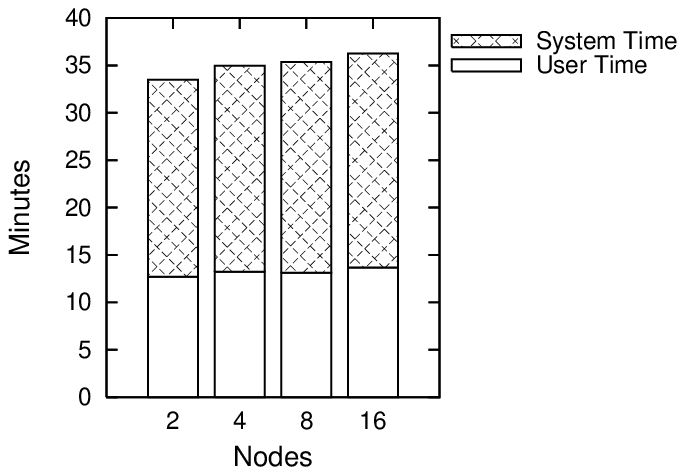}
		\label{fig:user_system_50}
	}
	\hfil	
	\subfigure[$M = 100$]{
		\includegraphics[width=2.3in]{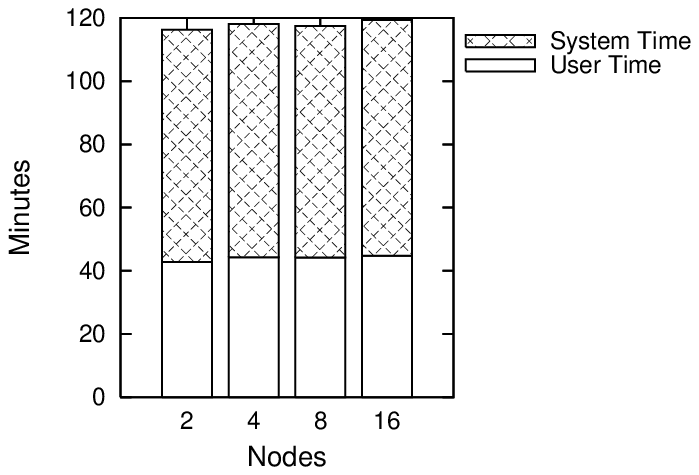}
		\label{fig:user_system_100}
	}
}
\caption{Individual user and system times with search depth $M$.  Note that the sum of system and user time does not equal actual execution time.  For example, a heavily loaded multiuser system may display the same system and user times as a fully dedicated system, but will have a longer real execution time due to the other running processes sharing the system.}
\label{fig:user_system}
\end{figure*}

Our first examination focused on the ratio of user to system time for the running program.  User time is time spent calculating, while system time is consumed by operating system overhead, such as memory allocation, file input and output, etc. \cite{biocomp06:Silberschatz9, biocomp06:Silberschatz12}.  This time breakdown can be obtained by using standard system libraries.  Figure \ref{fig:user_system} displays the composition of user and system time for search depths of 25, 50, and 100.  Immediately it can be seen that approximately $\frac{2}{3}$ of the execution time is spent on system overhead.  A look at the core calculation implementation illustrated in \ref{fig:calculate} explains where this extensive overhead originates.  Information is passed to each of the three subprograms (NewGenpp, svd, and mysolve) using a total of five files.  Each iteration performs five file reads and three file writes per data set.  Therefore a search depth of 100 (with two data sets) requires the system to handle 100,000 file reads and 60,000 file writes for each residue.

%--------------------------------------------------------------------------------%

\subsection{Removal of Intermediate Files}

\begin{table}
\renewcommand{\arraystretch}{1.3}
\caption{Actual execution time before and after removing intermediate files, using a fragment of length 14 running on 16 nodes.}
\label{tab:io_exec_time}
\centering
\begin{tabular}{|c||c||c|}
\hline
 & \multicolumn{2}{|c|}{\bfseries Execution Time (s)}\\
\hline
Depth $M$ & Intermediate Files & Internal Matrices \\
\hline\hline
25 & 59.4 & 13.2\\
\hline
50 & 224.6 & 35.0\\
\hline
100 & 648.6 & 53.9\\
\hline
\end{tabular}
\end{table}

The first step of optimization is to reduce system overhead by removing the intermediate files depicted in Figure \ref{fig:calculate}.  The files b.k.in and PolyGly.out.txt are still read in, but only once for each residue instead of for every iteration.  This is necessary because these files change depending on the current residue.  However, there are generally less than 100 residues being analyzed, making the overhead negligible.  Originally the three subprograms NewGenpp, svd, and mysolve were separate programs run using system calls, necessitating the use of intermediate files.  This also adds the overhead required for the system to create a new process for each call \cite{biocomp06:Silberschatz4}.  This was addressed by compiling all three programs into the main program, modified to take matrices as arguments instead of reading and writing files.  These three subprocesses are now accessed as function calls instead of system calls.

The implementation using internal data structures instead of intermediate files is much faster than the original, with sample runtimes documented in Table \ref{tab:io_exec_time}.  While actual runtimes can vary based on system load, these data were collected under similar conditions on an unloaded system and present a significant improvement.  With the system overhead in the core of the algorithm reduced, the algorithm as a whole can be evaluated to identify which sections can be improved.

%--------------------------------------------------------------------------------%

\subsection{Runtime Breakdown}

\begin{figure}
\centering
\includegraphics[width=1.2in]{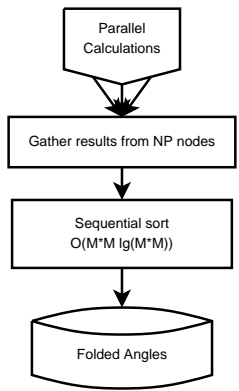}
\caption{Given a sorting algorithm with runtime $O(n lg n)$ and a maximum list length of $M^2$, the runtime for each sequential sort becomes $O(M^2 lg(M^2)$).}
\label{fig:sort}
\end{figure}

\begin{figure}
\centering
\includegraphics[width=3.4in]{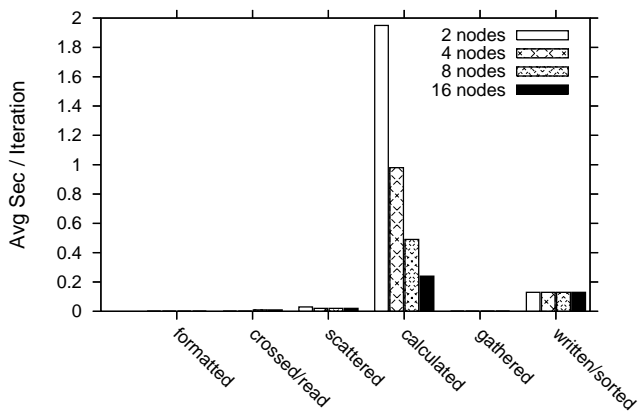}
\caption{Runtime breakdown after intermediate files have been eliminated, with a search depth $M$ of 100.  Note that the runtime consumed by the sequential sort is unaffected by the number of nodes.  Therefore, as the number of nodes increases the sort will begin to dominate runtime, and overall speedup will become sublinear.}
\label{fig:breakdown_noio_100}
\end{figure}

\begin{figure}
\centering
\includegraphics[width=3.4in]{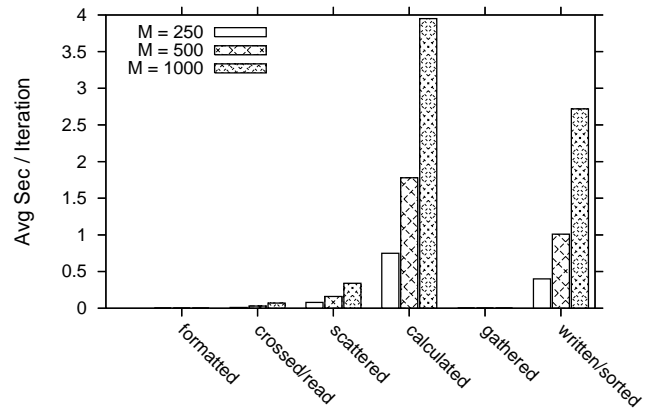}
\caption{Runtime breakdown after intermediate files have been eliminated, using 16 nodes with an increasing search depth $M$.  Here you can see the linear scaling of calculations and parallelization overhead (scattering and gathering), as well as the relative growth of the search.  Note that although the algorithm's complexity increases by a factor of $M^2$, in practice this factor approaches $M$ as $M$ increases.  This is due to each iteration combining a list of length $M$ with the new list of angles that often has length less than $M$.  In this example, these list lengths are generally less than 100. }
\label{fig:breakdown_noio_16n}
\end{figure}

The next step is to analyze the individual sections of the whole algorithm, in order to identify sections that can be improved.  Each section has a barrier after it, meaning that every node must complete a section before any node can procede to the next section.  Therefore if one node finishes the current section early, it must wait before continuing.  The six major sections identified are described as follows:

\begin{itemize}

\item \textbf{Formatted.}  The lists of angle combinations (stored in files) have been processed and trimmed to contain at most $M$ combinations.
 
\item \textbf{Crossed / Read.}  The lists of angles have been read into the program and all combinations have been formed in memory.

\item \textbf{Scattered.}  The combinations of angles that each node needs to calculate have been distributed to the respective nodes.  This section contributes to $OP$, or parallelization overhead.

\item \textbf{Calculated.}  All angle combinations for each node have been calculated.  This is done by the core of the program illustrated in Figure \ref{fig:calculate}.

\item \textbf{Gathered.}  The calculated fitness for each geometry has been collected from every node.  This section also contributes to the parallelization overhead $OP$.

\item \textbf{Written / Sorted.}  The list of all angle combinations and their calculated fitness has been written to a file.  The standard Linux \textit{sort} program is then called which sorts the file.  Note that this is done using only a single node, as illustrated in Figure \ref{fig:sort}.

\end{itemize}

Examples of this itemized analysis are shown in Figures \ref{fig:breakdown_noio_100} and \ref{fig:breakdown_noio_16n}.  For each of these figures, the maximum user time out of all nodes was recorded for each section.  These maximum values were then averaged for every residue iteration.  So for a run with 14 residues, 14 individual maximum values are recorded and averaged for each section.  Figure \ref{fig:breakdown_noio_100} illustrates the parallelization of the \textbf{calculated} section, and also the sequential runtime of the external sort.  As more nodes are added, this sorting begins to dominate total runtime and causes sublinear speedup relative to the number of nodes.

Figure \ref{fig:breakdown_noio_16n} shows the scalability relative to search depth $M$.  An interesting observation is that runtime does not reflect the quadratic relationship to $M$ that is shown in Equation \ref{eqn:complexity_stage2}.  Instead, runtime appears to increase linearly with respect to $M$.  This is due to each iteration combining a list of length $M$ with the new list of angles that often has length less than $M$.  The length $M_1$ of the combination list for previous residues achieves the maximum length $M$ after the first couple of iterations.  Search depth $M$ is only a maximum length, however, and the length $M_2$ of the angle list for the newly added residue is generally much smaller.  Equation \ref{eqn:complexity_stage2} can be broken down using $M_1$ and $M_2$ to become Equation \ref{eqn:complexity_stage2_breakdown}.  As search depth $M$ increases, $M_1$ becomes much larger than $M_2$, making runtime effectively linear with respect to $M$ given a fixed number of residues $N$ as shown in Equation \ref{eqn:complexity_stage2_linear}.

\begin{equation}
\label{eqn:complexity_stage2_breakdown}
O(N * M_1 * M_2)
\end{equation}

\begin{equation}
\label{eqn:complexity_stage2_linear}
O(N * M_1), M_1 \gg M_2
\end{equation}

%--------------------------------------------------------------------------------%

\subsection{Parallelization of Sort}

\begin{figure*}
\centering
\includegraphics[width=4.6in]{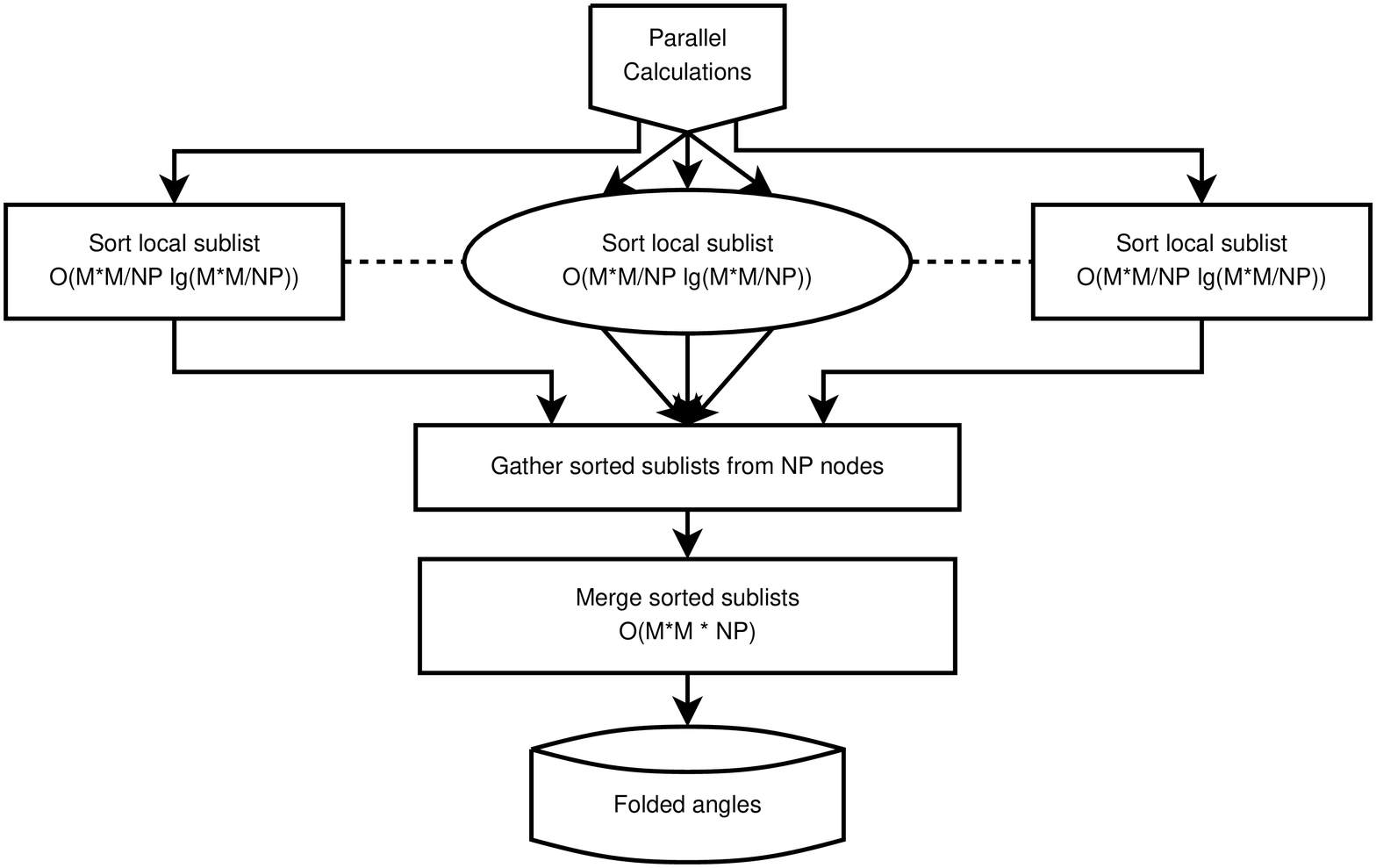}
\caption{By parallelizing the sort, each node has to sort $\frac{M^2}{NP}$ items instead of $M^2$.  Therefore each node sorts in $O(\frac{M^2}{NP} lg(\frac{M^2}{NP}))$ time.  With a final merge running in $O(M^2 * NP)$ time, total runtime for the parallel mergesort is given by Equation \ref{eqn:complexity_parallel_sort}.}
\label{fig:parallel_sort}
\end{figure*}

By calling an external sorting routine, the entire list of all combinations is sorted on the head node only.  This use of a nonparallel sorting algorithm reduces the scalability of the program, as shown in Figure \ref{fig:breakdown_noio_100}.  Recall that the runtime of Stage 2 with $N$ residues and a search depth of $M$ is quadratic with respect to $M$, as shown by Equation \ref{eqn:complexity_stage2}. However, sorting requires $O(n lg n)$ comparisons where $n$ is the length of the list to be sorted \cite{biocomp06:Cormen}.  Therefore the runtime of Stage 2 when considering the sorting section is given in Equation \ref{eqn:complexity_with_sort}, where the second term is the sequential sort as illustrated in Figure \ref{fig:sort}.  This means that as $M$ increases, the sort will begin to dominate runtime.

\begin{equation}
\label{eqn:complexity_with_sort}
O(N(M^2 + M^2 lg(M^2)))
\end{equation}

Although a sort requires $O(n lg n)$ runtime, a merge can be performed linearly in $O(n)$.  Therefore if each node sorted their sublist before sending it to the head node, the head node could then just merge the $NP$ sorted lists together.  A total of $NP - 1$ merges are required, producing $O(M^2 * NP)$ runtime since the maximum size of the list is $M^2$.  These merges can also be done in parallel, requiring only $lg(NP)$ steps.  The runtime for merging is negligible however, even on a single node.  Given a sorting runtime of $O(\frac{M^2}{NP} lg(\frac{M^2}{NP}))$ for each node (since their sublist is $\frac{M^2}{NP}$ items long and mergesort is $O(n lg n)$ \cite{biocomp06:Knuth}), the total runtime for the parallelized sort is shown in Equation \ref{eqn:complexity_parallel_sort}.  This sorting algorithm is illustrated in Figure \ref{fig:parallel_sort}.  In addition to improved asymptotic runtime of the parallel sort, the entire list of angle combinations with calculated fitness can be sorted in memory and does not have to be written to the disk every iteration.  While this only happens $N$ times, these files get quite large and outputting them takes a noticeable amount of time.

\begin{equation}
\label{eqn:complexity_parallel_sort}
O(N(\frac{M^2}{NP} lg(\frac{M^2}{NP}) + M^2 * NP))
\end{equation}

%%%%%%%%%%%%%%%%%%%%%%%%%%%%%%%%%%%%%%%%%%%%%%%%%%%%%%%%%%%%%%%%%%%%%%%%%%%%%%%%%%
%%%%%%%%%%%%%%%%%%%%%%%%%%%%%%%%%%%%%%%%%%%%%%%%%%%%%%%%%%%%%%%%%%%%%%%%%%%%%%%%%%

\section{Results}

\begin{figure}
\centering
\includegraphics[width=3.4in]{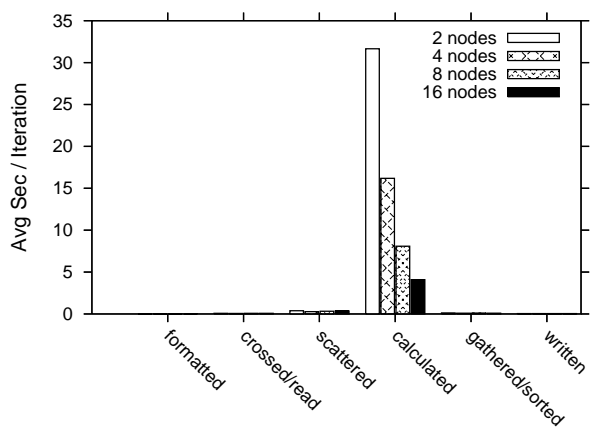}
\caption{Runtime breakdown using parallelized sort with a search depth $M$ of 1000.  Note that runtime dedicated to sorting has become negligible, and that parallelization overhead increases less for each node added.  This demonstrates the dominance of actual calculations to the overall runtime, and the desired linear scaling relative to the number of processors.}
\label{fig:breakdown_sort_1000}
\end{figure}

\begin{table}
\renewcommand{\arraystretch}{1.3}
\caption{Execution time of the optimized parallel algorithm using a fragment of length 14, running on 31 nodes.}
\label{tab:large_exec_time}
\centering
\begin{tabular}{|c||c|}
\hline
\bfseries Depth $M$ & \bfseries Execution Time (m)\\
\hline\hline
1000 & 3.6\\
\hline
2000 & 6.3\\
\hline
5000 & 17.2\\
\hline
\end{tabular}
\end{table}

Sample runtimes of {REDcRAFT} after optimized parallelization are given in Figure \ref{fig:breakdown_sort_1000}.  Note that runtime dedicated to sorting has become negligible with the new parallelism.  Although the parallelization overhead can be seen in the \textbf{scattered} section, it is also negligible compared to the performance increase given by the parallel \textbf{calculated} section.  Also, the increase in overhead for each additional node decays as the number of nodes increases, due to Equation \ref{eqn:communication_increase} which gives the amount of data transfer needed between nodes given a fixed search depth $M$.

\begin{equation}
\label{eqn:communication_increase}
\Theta(\frac{NP-1}{NP})
\end{equation}

Given the almost complete use of total runtime in the \textbf{calculated} section and the scalability of this section, the {REDcRAFT} algorithm as a whole becomes practically linearly scalable relative to the number of nodes.  As illustrated in Figure \ref{fig:breakdown_noio_16n}, the algorithm also approaches linearity relative to the search depth.  The runtimes of this optimized parallel algorithm are not even comparable to the original times, with some samples given in Table \ref{tab:large_exec_time}.  With a search depth of 1000 and fragment length of 14 the new algorithm finishes in under four minutes, and with a search depth of 5000 it finishes in under 20 minutes, while the initial implementation took hours to days.

%%%%%%%%%%%%%%%%%%%%%%%%%%%%%%%%%%%%%%%%%%%%%%%%%%%%%%%%%%%%%%%%%%%%%%%%%%%%%%%%%%
%%%%%%%%%%%%%%%%%%%%%%%%%%%%%%%%%%%%%%%%%%%%%%%%%%%%%%%%%%%%%%%%%%%%%%%%%%%%%%%%%%

\section{Discussion}

Continual departure from use of traditional types of experimental data and precipitously increasing reliance on RDC data necessitates the development of computational analysis methods of this rich source of structural data. In addition, considering the recent international efforts in increasing protein structure determination throughput, development of highly efficient and rapid methods of structure determination gains a new importance. {REDcRAFT} is designed specifically to address both needs. The linear scalability of {REDcRAFT} is a rare and yet highly desirable feature contributing to its utility. Considering today's cost of cluster computing environment, it is easy to envision the ability of converting a set of RDC data to a set of atomic coordinates in a matter of hours. 

The effects of efficiently parallelizing {REDcRAFT} transcend simply shortening the execution time. A more manageable computation time translates into the ability of simultaneous consideration of protein backbone structure, resonance assignment and characterization of internal motion. It is easy to envision a future version of {REDcRAFT} where an unassigned list of RDC data is easily converted to a list of atomic coordinates with full description of internal motion in an automated fashion.

%%%%%%%%%%%%%%%%%%%%%%%%%%%%%%%%%%%%%%%%%%%%%%%%%%%%%%%%%%%%%%%%%%%%%%%%%%%%%%%%%%
%%%%%%%%%%%%%%%%%%%%%%%%%%%%%%%%%%%%%%%%%%%%%%%%%%%%%%%%%%%%%%%%%%%%%%%%%%%%%%%%%%

%\newpage

\end{document}